\documentclass[aps,prl,twocolumn,groupedaddress,showpacs,showkeys]{revtex4}

\usepackage{amsmath}
\usepackage{mathrsfs}
\usepackage{amsfonts,amssymb}
\usepackage{graphicx}

\begin{document}

%\preprint{}

\title{Quantum Entanglement transfer between spin-pairs}

\author{Qing-You Meng}
%\email[Email:]{mengqyou@mail.nankai.edu.cn}
\affiliation{Theoretical Physics Division, Chern Institute of
Mathematics, Nankai University, Tianjin 300071, People's Republic of
China}

\author{Fu-Lin Zhang}
%\email[Email:]{flzhang@mail.nankai.edu.cn}
\affiliation{Theoretical Physics Division, Chern Institute of
Mathematics, Nankai University, Tianjin 300071, People's Republic of
China}

\author{Jing-Ling Chen}
\email[Email:]{chenjl@nankai.edu.cn}

\affiliation{Theoretical Physics Division, Chern Institute of
Mathematics, Nankai University, Tianjin 300071, People's Republic of
China}

\date{\today}

\begin{abstract}
We investigate the transfer of entanglement from source particles
(SP) to target particles (TP) in the Heisenberg interaction $ H=\vec
s_{1} \cdot \vec s_{2}$. In our research, TP are two qubits and SP
are two qubits or qutrits. When TP are two qubits, we find that no
matter what state the TP is initially prepared in, at the specific
time $t=\pi$, the entanglement of TP can attain to 1 after
interaction with SP which stay on the maximally entangled state. For
the  TP are two qutrits, we find that the maximal entanglement of TP
after interaction is relative to the initial state of TP and always
cannot attain to 1 to almost all of initial states of TP. But we
discuss an iterated operation which can make the TP to the maximal
entangled state.
\end{abstract}

\maketitle

\section{Introduction}

%\begin{figure}
%\includegraphics[width=0.45\textwidth]{fig1.jpg}
%\caption{\label{fig:epsart} The picture of the Model.}
%\end{figure}

Entanglement has been considered as a physical resource of quantum
information processing. It is profoundly important in quantum
teleportation [1], quantum computing [2], cryptography [3], and
quantum games [4,5]. Using Jayness-Cumming-type interaction between
atoms and cavity fields [6], entanglement transfer from two-mode
squeezed vacuum state to two separable atoms which are prepared in
pure state has been studied in Ref. [7]. Recently, Zou \emph{et al}
have discussed the entanglement transfer from some entangled
two-mode fields to mixed qubits [8] and Zhang \emph{et al} have
discussed the entanglement transfer from photons to atoms [9].

  In the paper [9], they conclude that the maximally entangled
state of photons can lead the atoms to the maximal state of
entanglement by nonlinear interaction. That is, the entanglement
among photons transfers to the atoms by the interaction. Similarly,
the transfer process between atoms and atoms stimulated our
interests and then we research it. Fortunately, we find some
interesting things in the cases of two pairs of qutrits. In the
process of study, we consider the Heisenberg interaction---$ H=\vec
s_{1} \cdot \vec s_{2}$, and mainly focus on the entanglement
transfer from source particles (SP) which are two qubits or qutrits
to the target particles (TP) which contain two qubits. When SP are
two qubits, the process of the transfer is clear and easy to
understand, because the quantity of entanglement between two qubits
is well-defined. But when SP are two qutrits, we have not a good
definition of entanglement degree and only have two invariants
$I_{1}$ and $I_{2}$ [10]. In the second section, we introduce the
model and the study method. In the third section, the transfer
between two pairs of qubits is studied. And in the Fourth section,
we study the cases of transfer from two qutrits to two qubits. The
Final section is the conclusion.

\section{The Model and The Research Method}

\begin{figure}
\includegraphics[width=0.45\textwidth]{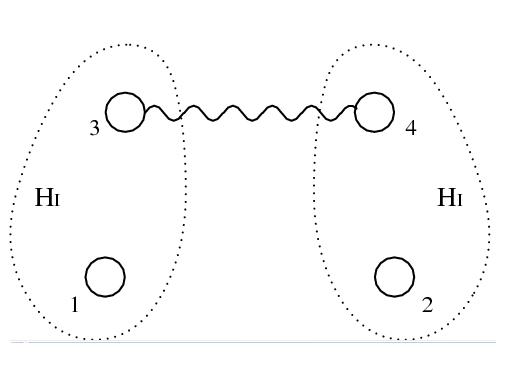}
\caption{\label{fig:epsart} The picture of the Model.}
\end{figure}

 As shown in Fig.1, there are four particles in this model, and two of them 1 and 2 are called
target particles (TP) which are two qubits and initially stay on the
pure state $|\psi_{12}\rangle$
\begin{eqnarray}\label{1}
|\psi_{12}\rangle=\cos \theta_{1} |00\rangle+\sin \theta_{1}
|11\rangle
\end{eqnarray}

The other two particles 3 and 4 are called source particles (SP)
which are two qubits or two qutrits and initially stay on the pure
state $|\psi_{34}\rangle$. The initial density operator of the whole
system is

\begin{eqnarray}\label{2}
\rho(0)=\rho_{12}(0)\otimes \rho_{34}(0)=|\psi_{12}\rangle \langle
\psi_{12}|\otimes|\psi_{34}\rangle \langle \psi_{34}|
\end{eqnarray}
where $\rho_{12}(0)$ is the initial density operator of TP and
$\rho_{34}(0)$ is similar. Then, particles 1 and 3 interact under
the interaction Hamiltonian $H=\vec s_{1} \cdot \vec s_{2}$ and
particles 2 and 4 are also acted with H.

  The time evolution operator of the whole system can be derived $U(t)=u_{13}(t)\otimes
u_{24}(t)$, where $u_{13}(t)$ is the time operator between particles
1 and 3, then $u_{24}(t)$ is similar. Now we can obtain the density
operator $\rho(t)$ of the system at time t

\begin{eqnarray}\label{3}
\rho(t)&=&U(t)\rho(0)U^{\dag}(t)\nonumber\\
&=&u_{13}(t)\otimes u_{24}(t)\rho_{12}(0)\otimes
\rho_{34}(0)u_{13}^{\dag}(t)\otimes u_{24}^{\dag}(t)
\end{eqnarray}

To get the entanglement between two target particles, one needs to
trace over the SP 3 and 4 to obtain the reduced density operator for
the TP $\rho_{12}(t)$

\begin{eqnarray}\label{4}
\rho_{12}(t)=\emph{\emph{Tr}}_{3,4}\rho(t)
\end{eqnarray}

From (1), we can calculate the initial quantity of entanglement
$E_{12}(0)$ of TP
\begin{eqnarray}\label{5}
E_{12}(0)=2|\sin \theta_{1}\cos \theta_{1} |
\end{eqnarray}
where we use the Negativity [11,12] to denote the quantity of
entanglement. For a bipartite system described by the density
matrix, the negativity criterion for entanglement of the two
subsystems is given by the following formula: $\varepsilon
=-2\sum_{i} \lambda_{i}^{-} $ where the sum is taken over the
negative eigenvalues $\lambda_{i}^{-}$ of the partial transposition
of the density matrix $\rho$.The value $\varepsilon=0$ indicates
that the two subsystems are separable. This function varies between
0 and 1, and monotonically increases as the entanglement grows. Then
from (4), we can calculate the Negativity of TP at the time
t---$E_{12}(t)$. We can know the detailed process of the
entanglement transfer from SP to TP by studying the $E_{12}(t)$.

\section{Entanglement transfer from two qubits to two qubits}

The initial state of SP $|\psi_{34}\rangle$ is

\begin{eqnarray}\label{6}
|\psi_{34}\rangle=\cos \theta_{2} |00\rangle+\sin \theta_{2}
|11\rangle
\end{eqnarray}

And the interaction Hamiltonian between 1 and 3 is

\begin{eqnarray}\label{7}
H=\vec s_{1} \cdot \vec
s_{3}=\frac{1}{4}\left(\begin{array}{cccc}1&0&
0&0\\
0&-1&2&0\\
0&2&-1&0\\0&0&0&1
\end{array}\right)
\end{eqnarray}

 The time evolution operator between
particle 1 and 3 $u_{13}(t)$ is

\begin{eqnarray}\label{8}
u_{13}(t)=e^{-\frac{i t}{4}}\left(\begin{array}{cccc}1&0&
0&0\\
0&\frac{1}{2}(1+e^{i t})&-\frac{1}{2}(-1+e^{i t})&0\\
0&-\frac{1}{2}(-1+e^{i t})&\frac{1}{2}(1+e^{i t})&0\\0&0&0&1
\end{array}\right)
\end{eqnarray}

The interaction Hamiltonian and the time evolution operator between
particle 2 and 4 are similar with particle 1 and 3.

The initial quantity of entanglement of SP $E_{34}(0)$ is

\begin{eqnarray}\label{9}
E_{34}(0)=2|\sin \theta_{2}\cos \theta_{2} |
\end{eqnarray}

After some calculations, the reduced density operator $\rho_{12}(t)$
for TP is derived

\begin{eqnarray}\label{10}
\rho_{12}(t)=\left(\begin{array}{cccc}A(t)&0&
0&F(t)\\
0&B(t)&0&0\\
0&0&C(t)&0\\F^{\ast}(t)&0&0&D(t)
\end{array}\right)
\end{eqnarray}

where the matrix basis is chosen as $\{|00\rangle, |01\rangle,
|10\rangle,|11\rangle\}$. The coefficients in Eq. (10) are functions
of time and given by

\begin{eqnarray}\label{11}
A(t)&=&(\cos \theta_{2}\sin \theta_{1}\sin^{2}{\frac{t}{2}}-\sin
\theta_{2}\cos
\theta_{1}\cos^{2}{\frac{t}{2}})^{2}\nonumber\\&&+\cos^{2} \theta_{1}\cos^{2} \theta_{2} \nonumber\\
 B(t)&=&\frac{1}{4}\sin^{2}{t}\sin^{2}
(\theta_{1}+ \theta_{2})\nonumber\\
 C(t)&=& B(t)\nonumber\\
  D(t)&=&(\cos \theta_{2}\sin
\theta_{1}\cos^{2}{\frac{t}{2}}-\sin \theta_{2}\cos
\theta_{1}\sin^{2}{\frac{t}{2}})^{2}\nonumber\\&&+\sin^{2}
\theta_{1}\sin^{2} \theta_{2}\nonumber\\
 F(t)&=&e^{-i t}[-\cos
\theta_{2}\sin \theta_{2}\sin^{2}{\frac{t}{2}}(\cos^{2}
\theta_{1}+e^{2i t}\sin^{2} \theta_{1})\nonumber\\&&+\cos
\theta_{1}\sin \theta_{1}\cos^{2}{\frac{t}{2}}(\cos^{2}
\theta_{2}+e^{2i t}\sin^{2} \theta_{2})]
\end{eqnarray}

And the analytic expression of the quantity of entanglement of TP at
the time t, $E_{12}(t)$ is very complicated. And one who wants to
know its real form, can calculate it using the following
formula---$E_{12}(t)=\emph{\emph{Max }}\{0,
\sqrt{(B-C)^2+4|F|^2}-(B+C)\}$. From now on, we use the numerical
method to study the problem and give some numerical results.

  First, we find that $E_{12}(t)$ has a period that is $2\pi$, that
  is
to say
\begin{eqnarray}\label{12}
E_{12}(2k\pi)=E_{12}(0)=2|\sin \theta_{1}\cos \theta_{1} |
\end{eqnarray}
where $k$ is the natural number.

  Second, when $t=\pi$,
\begin{eqnarray}\label{13}
E_{12}(\pi)=E_{34}(0)=2|\sin \theta_{2}\cos \theta_{2} |
\end{eqnarray}
That is to say, the entanglement of TP can attain to the
entanglement of SP. Assuming that the entanglement of SP is 1 and
initially the entanglement of TP is very small, we can enhance the
entanglement of TP and make them stay on the maximal entangled state
under the interaction $H$.

  Finally, we give some numerical pictures to show the propriety of
the cases.
\begin{figure}
\includegraphics[width=0.5\textwidth]{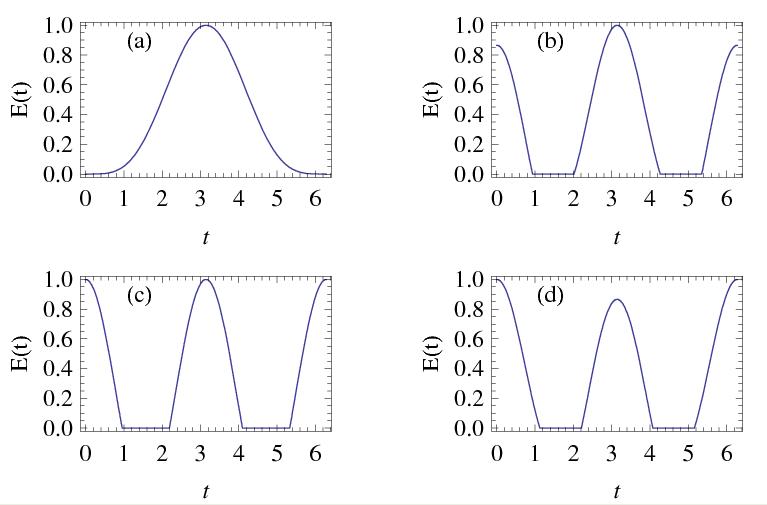}
\caption{\label{fig:epsart} The entanglement of 1,2 particles
$E_{12}$ as a function of $t$. (a).$\theta_{1}=0 ,
\theta_{2}=\frac{\pi}{4}$; (b).$\theta_{1}=\frac{\pi}{6} ,
\theta_{2}=\frac{\pi}{4}$; (c).$\theta_{1}=\frac{\pi}{4} ,
\theta_{2}=\frac{\pi}{4}$; (d).$\theta_{1}=\frac{\pi}{4} ,
\theta_{2}=\frac{\pi}{6}$.}
\end{figure}

From the Fig.2, one can see that when $E_{12}(0)=0, E_{34}(0)=1$
 and as time goes by, the entanglement of TP increases from 0 to 1
when $t=\pi$; then decreases from 1 to 0. To the picture (b), (c)
and (d), the initial entanglement $E_{12}(0)$ of TP is not equal to
0. The process of transfer contains five sections. (\romannumeral1).
the entanglement of TP $E_{12}$ decreases from $E_{12}(0)$ to zero;
(\romannumeral2). $E_{12}$ stays on the zero during a period of
time; (\romannumeral3). $E_{12}$ increases from zero to $E_{34}(0)$
when $t=\pi$ and then decreases from $E_{34}(0)$ to zero;
(\romannumeral4). $E_{12}$ also stays on the zero during a period of
time; (\romannumeral5). $E_{12}$ increases from zero to $E_{12}(0)$
when $t=2\pi$.

\section{Entanglement transfer from two qutrits to two qubits}

In this section, the SP are two qutrits and so we research the
transfer process from two qutrits to two qubits. The initial state
of SP $|\psi_{34}\rangle$ is
\begin{eqnarray}\label{14}
|\psi_{34}\rangle=k_{0} |00\rangle+k_{1} |11\rangle+k_{2} |22\rangle
\end{eqnarray}
where $k_{0},k_{1},k_{2}$ satisfy the normalizing condition
$|k_{0}|^{2}+|k_{1}|^{2}+|k_{2}|^{2}=1$.

 And the interaction Hamiltonian between 1 and 3 is

\begin{eqnarray}\label{15}
H=\vec s_{1} \cdot \vec
s_{3}=\left(\begin{array}{cccccc}\frac{1}{2}&0&
0&0&0&0\\
0&0&0&\frac{1}{\sqrt{2}}&0&0\\
0&0&-\frac{1}{2}&0&\frac{1}{\sqrt{2}}&0\\0&\frac{1}{\sqrt{2}}&0&-\frac{1}{2}&0&0\\0&0&\frac{1}{\sqrt{2}}&0&0&0\\0&
0&0&0&0&\frac{1}{2}\\
\end{array}\right)
\end{eqnarray}

The time evolution operator between particle 1 and 3 $u_{13}(t)$ is
\begin{eqnarray}\label{16}
u_{13}(t)=\left(\begin{array}{cccccc}x_1&0&
0&0&0&0\\
0&x_2&0&x_3&0&0\\
0&0&x_4&0&x_3&0\\0&x_3&0&x_4&0&0\\0&0&x_3&0&x_2&0\\0&
0&0&0&0&x_1\\
\end{array}\right)
\end{eqnarray}
where $x_1=e^{-\frac{i t}{2}},x_2=\frac{e^{i t}+2e^{-\frac{i
t}{2}}}{3},x_3=-\sqrt{2}\frac{-e^{i t}+e^{-\frac{i
t}{2}}}{3},x_4=\frac{2e^{i t}+e^{-\frac{i t}{2}}}{3}$.

Just as the last section, particle 2 and 4 do the same operation
with particle 1 and 3 in the interaction Hamiltonian and time
evolution operator.
  The two invariants of $|\psi_{34}\rangle$ is [10]
\begin{eqnarray}\label{17}
I_1 & = &|k_{0}|^{4}+|k_{1}|^{4}+|k_{2}|^{4} \nonumber\\
I_2 & = &|k_{0}|^{6}+|k_{1}|^{6}+|k_{2}|^{6}
\end{eqnarray}

 After some calculations, the reduced density operator $\rho_{12}(t)$ for TP is derived
\begin{eqnarray}\label{18}
\rho_{12}(t)=\left(\begin{array}{cccc}a(t)&0&
0&e(t)\\
0&b(t)&0&0\\
0&0&c(t)&0\\e^{\ast}(t)&0&0&d(t)
\end{array}\right)
\end{eqnarray}

where the matrix basis is chosen as $\{|00\rangle, |01\rangle,
|10\rangle,|11\rangle\}$. The coefficients in Eq. (18) are functions
of time and given by

\begin{widetext}
\begin{eqnarray}\label{19}
a(t)&=&k_{0}^2 \cos^{2} \theta_{1} +
 \frac{e^{-3 i t} }{81}[(2 + e^{\frac{3 i t}{2}})^2 k_{1} \cos \theta_{1} +2 (-1 + e^{\frac{3 i t}{2}})^2 k_{0} \sin \theta_{1}] [(1 +
       2 e^{\frac{3 i t}{2}})^{2} k_{1} \cos \theta_{1} +2 (-1 + e^{\frac{3 i t}{2}})^2 k_{0} \sin \theta_{1}] +\nonumber\\
    &&
 \frac{e^{-3 i t}}{81} [(2 + e^{\frac{3 i t}{2}})^2 k_{2} \cos \theta_{1} +
    2 (-1 + e^{\frac{3 i t}{2}})^2 k_{1} \sin \theta_{1}][(1 +
       2 e^{\frac{3 i t}{2}})^{2} k_{2} \cos \theta_{1} +
    2 (-1 + e^{\frac{3 i t}{2}})^2 k_{1} \sin
    \theta_{1}]\nonumber\\
b(t)&=&-\frac{2 e^{-3 i t}}{81}(-1 + e^{\frac{3 i t}{2}})^2 [(1+ 2
e^{\frac{3 i t}{2}})^{2} k_{1} \cos \theta_{1} + (2 + e^{\frac{3 i
t}{2}})^2 k_{0} \sin \theta_{1}] [(2 + e^{\frac{3 i t}{2}})^{2}
k_{1} \cos \theta_{1} +(1 +
       2 e^{\frac{3 i t}{2}})^2 k_{0} \sin \theta_{1}] \nonumber\\
    &&-
 \frac{2e^{-3 i t}}{81}(-1 + e^{\frac{3 i t}{2}})^2 [(1 + 2 e^{\frac{3 i t}{2}})^{2} k_{2} \cos \theta_{1}
 + (2 + e^{\frac{3 i t}{2}})^2 k_{1} \sin \theta_{1}] [(2 + e^{\frac{3 i t}{2}})^{2} k_{2} \cos \theta_{1} +
   (1 + 2 e^{\frac{3 i t}{2}})^2 k_{0} \sin \theta_{1}]\nonumber\\
c(t)&=& b(t)\nonumber\\
d(t)&=&k_{2}^2 \sin^{2} \theta_{1}+
 \frac{e^{-3 i t}}{81}[2 (-1 + e^{\frac{3 i t}{2}})^2 k_{1} \cos \theta_{1} +(2 +
       e^{\frac{3 i t}{2}})^2 k_{0} \sin \theta_{1}] [2 (-1 + e^{\frac{3 i t}{2}})^2 k_{1} \cos \theta_{1} + (1 +
       2 e^{\frac{3 i t}{2}})^2 k_{0} \sin \theta_{1}] +\nonumber\\
    &&
 \frac{e^{-3 i t}}{81}  [2 (-1 + e^{\frac{3 i t}{2}})^2 k_{2} \cos \theta_{1} + (2 +
       e^{\frac{3 i t}{2}})^2 k_{1} \sin \theta_{1}][2 (-1 + e^{\frac{3 i t}{2}})^2 k_{2} \cos \theta_{1} + (1 +
       2 e^{\frac{3 i t}{2}})^2 k_{1} \sin \theta_{1}]\nonumber\\
e(t)&=& \frac{e^{-3 i t}}{9}
   k_{0} \cos \theta_{1} [2 (-1 + e^{\frac{3 i t}{2}})^2 k_{1} \cos \theta_{1} + (2 +
       e^{\frac{3 i t}{2}})^2 k_{0} \sin \theta_{1}] \nonumber\\
    &&+
 \frac{1}{9} k_{2} \sin \theta_{1} [(1 + 2 e^{\frac{3 i t}{2}})^2 k_{2} \cos \theta_{1} +
    2 (-1 + e^{\frac{3 i t}{2}})^2 k_{1} \sin \theta_{1}] \nonumber\\
    &&+
 \frac{e^{-3 i t}}{81}  [(2 + e^{\frac{3 i t}{2}})^2 k_{1} \cos \theta_{1} +
    2 (-1 + e^{\frac{3 i t}{2}})^2 k_{0} \sin \theta_{1}] [2 (-1 + e^{\frac{3 i t}{2}})^2 k_{2} \cos \theta_{1} + (1 +
       2 e^{\frac{3 i t}{2}})^2 k_{1} \sin \theta_{1}]
\end{eqnarray}
\end{widetext}
Compared with the last section, the analytic expression of the
quantity of entanglement of TP at the time t---$E_{12}(t)$ is more
complicated. We still use the numerical method to study the problem
and give some numerical results from now on.

  First, we find that $E_{12}(t)$ has a period that is $\frac{4\pi}{3}$
\begin{eqnarray}\label{20}
E_{12}\left(\frac{4k\pi}{3}\right)=E_{12}(0)=2|\sin \theta_{1}\cos
\theta_{1} |
\end{eqnarray}
where $k$ is the natural number.

Second, being similar to the case that SP are two qubits, when
$t=T/2$ i.e. $t=2\pi/3$, SP can influence the entanglement of TP to
a great extent. So we can calculate the analytic expression of the
reduced density operator $\rho_{12}(t)$ and the entanglement at the
time $t=\frac{2\pi}{3}$, denoted as $\rho_{12}(\frac{2\pi}{3})$,
$E_{12}(\frac{2\pi}{3})$. Obviously, $E_{12}(\frac{2\pi}{3})$ is a
function of $\{\theta_{1},k_{0},k_{1},k_{2}\}$ and by some
transformations the parameters can become
$\{\theta_{1},I_{1},I_{2}\}$. For a given value of $\theta_{1}$,
there is a pair of $I_{1}, I_{2}$ that makes the entanglement
$E_{12}(\frac{2\pi}{3})$ to the maximal quantity. Considering the
complexity of analytical form of $E_{12}(\frac{2\pi}{3})$, we have
to use the numerical method to find the maximal quantity of
$E_{12}(\frac{2\pi}{3})$, denoted as $E_{12Max}(\frac{2\pi}{3})$,
and $I_{1}, I_{2}$ matched with $E_{12Max}(\frac{2\pi}{3})$ to a
special $\theta_{1}$.

\begin{figure}
\includegraphics[width=0.5\textwidth]{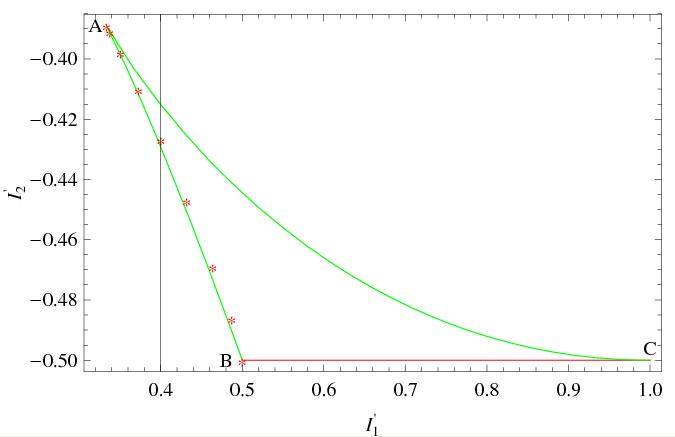}
\caption{\label{fig:epsart} The picture of $I_{2}^{'}-I_{1}^{'}$ and
9 maximal point matched with $\theta_{1}=0, \frac{\pi}{32},
\frac{\pi}{16}, \frac{3\pi}{32}, \frac{\pi}{8}, \frac{5\pi}{32},
\frac{3\pi}{16}, \frac{7\pi}{32}, \frac{\pi}{4}$.}
\end{figure}

  Now, we introduce how to find the $E_{12Max}(\frac{2\pi}{3})$. Firstly, the area to satisfy the
physical condition is very small in the picture of number pair
$I_{1}-I_{2}$. To show it clearly, we need to make a transformation
\begin{eqnarray}\label{21}
I_1^{'} & = &I_1  \nonumber\\
I_2^{'} & = &I_2-\frac{3}{2}I_1
\end{eqnarray}
The physical region transform in the Fig. 3 and A, B, C three points
are:\\

\begin{tabular}{|c|c|c|c|}
\hline Points&
$I_{1}$&$I_{2}$&$(k_{0},k_{1},k_{2})$\\
\hline A&$\frac{1}{3}$&$\frac{1}{9}$& $(\frac{\sqrt{3}}{3},\frac{\sqrt{3}}{3},\frac{\sqrt{3}}{3})$\\
\hline
B&$\frac{1}{2}$&$\frac{1}{4}$&$(\frac{\sqrt{2}}{2},\frac{\sqrt{2}}{2},0),(\frac{\sqrt{2}}{2},0,\frac{\sqrt{2}}{2}),(0,\frac{\sqrt{2}}{2},\frac{\sqrt{2}}{2})$\\
\hline C&1&1& $(1,0,0),(0,1,0),(0,0,1)$\\\hline
\end{tabular}
\\\\

From Fig. 3   one can see the 9 maximal points are almost on the
line AB. The real equation of frontier AB is very complicated, so we
give an approximate one. That is
\begin{eqnarray}\label{22}
I_2^{'} & = &-\frac{2}{3}I_1^{'}-\frac{1}{6}
\end{eqnarray}

And the error is small or equal to $1\%$. So we can infer that to a
special $\theta_{1}$, the point---$\{I_{1},I_{2}\}$ that makes
$E_{12}(\frac{2\pi}{3})$ attain to the maximal quantity is
approximately on the line AB. Now, you may ask that how $\theta_{1}$
is matched with $I_{1},I_{2}$. By some calculations, we can get a
better approximate result
\begin{eqnarray}\label{23}
I_1 & = &-0.08756\sin^2 2\theta_1-0.07911\sin 2\theta_1+0.5
\end{eqnarray}
Its error is small or equal to $1.4\%$. Now, if you give a
$\theta_{1}$, you can first use Eq.(23) to get $I_1$ and then lead
the result you calculated to the Eq.(22) and (21), then you get the
$I_{1}, I_{2}$ which matched with the maximal
$E_{12Max}(\frac{2\pi}{3})$. And the value of
$E_{12Max}(\frac{2\pi}{3})$ can be obtained.

Secondly, from the above table, we know point A and B are very
special in all of the states. State A is the maximal entangled state
of two qutrits. And the form of state B is likely with the maximal
entangled state of two qubits. But the state B cannot elevate the
entanglement of TP to unity and only to 0.8.

Now, some interesting things happened. To the case that SP are two
qubits, the entanglement of TP after interaction must attain to 1
when SP are in the maximally entangled state. But the case that SP
are two qutrits is very different from two qubits. The maximally
entangled state of the qutrits cannot make all states of TP attain
to the maximal state as the Fig.4. shown.

More interesting is that there does not exist a state that can make
the entanglement of state of TP to unity in all states of qutrits.
That means that there is a special state matched with a special
$\theta_{1}$ that can make the entanglement of states of TP to a
maximal quantity which less than 1. The details have been shown in
Fig. 4.
\begin{figure}
\includegraphics[width=0.5\textwidth]{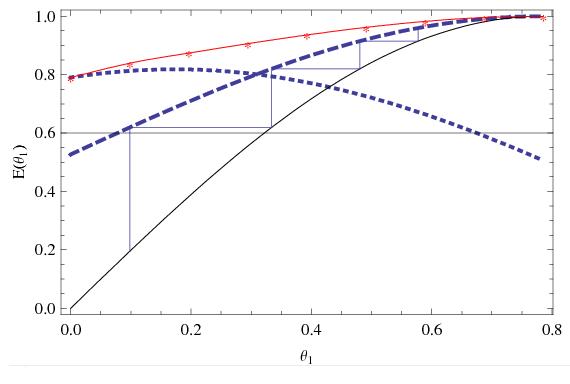}
\caption{\label{fig:epsart} Black real line denotes the initial
entanglement of TP; dashed line denotes the entanglement of TP when
$t=\frac{2\pi}{3}$ after interaction with SP which stay on the A
state initially; dotted line denotes the entanglement of TP when
$t=\frac{2\pi}{3}$ after interaction with SP which stay on the B
state initially; and the red real line denotes the maximal
entanglement quantity of TP in all states of qutrits after
interaction; the fold line denotes the process that entanglement of
TP initially is 0.2 and attains to 0.96 after interaction with SP 4
times.}
\end{figure}

From the Fig. 4, we can see that all states of qutrits cannot
elevate the entanglement of TP to 1 except the case that when SP is
on the state A and the initial state of TP is maximally entangled
state. But, if you want to make the entanglement of TP attain to 1
only using state A, you can make many times just like the fold line
in Fig.4 has shown. It is after interacting with SP which stay on
state A that the entanglement of TP is from initial 0.2 to 0.62.
Similarly, with 0.62 as the initial value, the entanglement of TP is
from 0.62 to 0.82 after interacting with SP which stay on state A
again. Doing this operation iteratively, after interacting with SP
four times, the entanglement of TP can attain to 0.96.

\section{Discussion and Conclusions}

In the interaction Hamiltonian $ H=\vec s_{1} \cdot \vec s_{2}$, the
entanglement transfer from SP to TP is investiged. We mainly focus
on two cases.

Case 1: When SP are two qubits, we can enhance the entanglement of
TP to 1 by interacting with SP whose entanglement is 1.

Case 2: When SP are two qutrits, generally, we cannot enhance the
entanglement of TP to 1 by the interaction with SP, even if SP is on
the maximally entangled state. More interesting thing is that to a
special initial entanglement of TP, there exists a state in all
states of qutrits which satisfy the physical condition. Compared
with other states of qutrits, this state can make a largest
enhancement to the entanglement of TP after interaction. Generally,
this state is not the maximally entangled state, but it can make TP
to a larger entangled state than the maximally entangled state did.

Considering this thing carefully, we find that in the case 1, the
maximal entanglement of TP after interaction has no relation with
the initial state of TP and always can attain to 1 to all initial
states of TP, as long as SP stay on the maximally entangled state
initially. In the case 2, the maximal entanglement of TP after
interaction is relative to the initial state of TP and always cannot
attain to 1 to all states of TP except one condition that the
initial state of TP and SP is the maximally entangled states. But
this case is useless. Because the initial state of TP is already the
maximally entangled state, it is not necessary to interact with SP.
Therefore, we only have interest on the initial state of TP which is
not maximally entangled state.

From the above discussion, we find that there are some differences
between case 1 and case 2. Firstly, the relation between  the
maximal entanglement of TP after interaction and the entanglement of
TP initially is different. Secondly, the behavior and function of
maximally entangled state is different in the entanglement transfer
process. Our next work is to give the reason that brings about those
differences exactly.

Now, we consider the reason from two aspects. On the one hand, the
difference derives from the form of interaction. So we will use more
general form of interaction, for example $H=J_1 s_1 ^{x}s_2 ^{x}+J_2
s_1 ^{y}s_2 ^{y}+J_3 s_1 ^{z}s_2 ^{z}+B_1 s_1 ^{z}+B_2 s_2 ^{z}$.
And we will adjust the quantity of those parameters and see whether
the difference will vanish or not. On the other hand, we conjecture
that the difference of entanglement between qubits and qutrits lead
to those differences. As we all know, there does not exist a well
concept to describe the entanglement of two qutrits. We expect to
find a proper and general definition of entanglement of two qutrits
based on the entanglement transfer to obtain the essential reason.

This work is supported in part by NSF of China Grant Nos. 10575053
and 10605013 , Program for New Century Excellent Talents in
University, and the Project-sponsored by SRF for ROCS, SEM.

\end{document}